\documentclass[a4paper,11pt,notc]{article}
\usepackage{jheppub,booktabs}
\usepackage{amsmath,amssymb}
\usepackage{cancel}
\usepackage{mathrsfs}
\usepackage{graphicx}
\usepackage{hepunits}
\usepackage{wasysym}
\usepackage{feynmf}
\usepackage[utf8x]{inputenc}

\def\bm#1{\mbox{\boldmath$#1$\unboldmath}}

\author{Florian Goertz}
\affiliation{
Institute for Theoretical Physics, \\
ETH Zurich, 8093 Zurich, Switzerland}
\emailAdd{fgoertz@itp.phys.ethz.ch}
\title{An Indirect Handle on the Down-Quark Yukawa Coupling}
\abstract{To measure the Yukawa couplings of the up and down quarks, $Y_{u,d}$, seems to be far beyond the capabilities
of current and (near) future experiments in particle physics.
By performing a general analysis of the potential misalignment between quark masses and Yukawa couplings,
we derive predictions for the magnitude of induced flavor-changing neutral currents (FCNCs), depending on the shift in 
the physical Yukawa coupling of first-generation quarks. We find that a shift of $-100\%$ in the down-quark Yukawa 
$Y_d$ would generically result in $ds$ transitions in conflict with Kaon physics. This could already be seen as 
evidence for a non-vanishing direct coupling of the down quark to the newly discovered Higgs boson. The {\it non-observation}
of certain, already well-constrained, processes is thus turned into a powerful indirect measure of physical parameters of 
the effective Standard-Model, which are so far basically unconstrained from experiment and extremely challenging 
to access with other methods. In particular, we can already deduce that $Y_d$ should vary at most by $\sim 50\%$ 
from its Standard Model value, barring an alignment of new physics effects with the SM
Yukawa couplings. Such an (orthogonal) alignment scenario is however in general much easier to test at the LHC.
Similarly, improvements in limits on FCNCs in the up-type quark sector can lead to valuable information on the physical
Yukawa coupling of the up-quark.
}
\date{\today}
\notoc
\begin{document}

\maketitle
\newpage
\section{Introduction}
While the LHC and a future linear collider are expected to determine the Yukawa couplings of the heavy third-generation 
fermions at the ${\cal O}$(\%) level, see e.g. \cite{Peskin:2012we,Peskin:2013xra}, the diagonal Yukawa couplings
of the first generation seem to be out of our direct reach in the near future. In this paper, we will demonstrate how we
can nevertheless gain valuable information about the possible size of the up and in particular the down-quark yukawa 
couplings in the mass basis, $Y_{u,d}$, in an indirect way. For that purpose we employ the fact that modifications of the Standard-Model (SM)
Yukawa matrices that change $Y_d$ generically also induce modifications in the off-diagonal entries in the mass basis 
and thus lead to flavor-changing neutral currents (FCNCs). In particular, tree-level Higgs exchange can now mediate 
meson-antimeson oscillations. These are however severely constrained from flavor-physics experiments, see 
e.g.~\cite{Bona:2007vi}.
 
\section{Setup}
We want to examine a possible misalignment between the quark-mass and Yukawa matrices. In order to keep the discussion 
general and to respect gauge invariance, we introduce this misalignment via the $D=6$ operators	\begin{equation}
	\label{eq:D6}
	{\cal L}_6^Y =\frac{1}{v^2}\left( (\Phi^\dagger \Phi)\,\bar q_L \bm{C}_u \Phi^c u_R +  (\Phi^\dagger \Phi)\, \bar q_L 
	\bm{C}_d \Phi\,d_R\right)\,.
	\end{equation}
Here, $\Phi$ denotes the Higgs doublet, which we will parametrize in unitary gauge as $\Phi=1/\sqrt 2 \left(0,h+v\right)^T$,
where $v$ is the vacuum expectation value $\langle \Phi \rangle = 1/\sqrt 2\,\left(0,v\right)^T $, $h$ is the physical
Higgs field, and $q_L, u_R, d_R$ are the chiral SM-quark doublet and singlets, each 3-vectors in flavor space. Inserting this 
decomposition of the Higgs doublet into (\ref{eq:D6}) as well as into the SM Yukawa terms with couplings 
$\hat{\bm{Y}}^{u,d}_{\rm SM}$, we arrive at the fermion masses and Higgs couplings in the flavor basis
	\begin{equation}
	\label{eq:LM}
	{\cal L} \supset - u_L \left( \hat{\bm{M}}^u + \frac{1}{\sqrt 2} \hat{\bm{Y}}^u\, h \right) u_R 
	-d_L \left( \hat{\bm{M}}^d +  \frac{1}{\sqrt 2} \hat{\bm{Y}}^d\, h \right)  d_R\, ,
	\end{equation}
where $\hat{\bm{Y}}^{u,d} =\hat{\bm{Y}}^{u,d}_{\rm SM} + \frac 3 2\, \bm{C}_{u,d}$ and 
$\hat{\bm{M}}^{u,d} =\frac{v}{\sqrt 2}(\hat{\bm{Y}}^{u,d}_{\rm SM} + \frac 1 2\bm{C}_{u,d})= \frac{v}{\sqrt 2}
( \hat{\bm{Y}}^{u,d} -  \bm{C}_{u,d})$ are now independent parameters. 

In this article, we focus on the light quarks and compare our predictions with the strongest constraints 
on corresponding quark FCNCs available to date. The only ``assumption'' we make is that the Wilson coefficients 
$\bm{C}_{u,d}$ exhibit an anarchic flavor structure, in a sense that they feature arbitrary complex entries, up to a certain
allowed scale. While a full comprehensive channel-by-channel analysis, including a detailed survey 
of the impact of further assumptions we could make on the structure of the dimension-6 operators (\ref{eq:D6}), as well 
as an extension to the lepton (and heavy quark) sector, would be interesting, here we just want to point out the predictive 
power of the approach. A corresponding global analysis is left for future work \cite{GinPrep}. 

Let us nevertheless already note that abandoning the chosen anarchic approach by assuming a special flavor structure like
an (approximate) alignment of  the $D=6$ operators with the Yukawa matrices, $\hat{\bm{M}}^{u,d} \propto \hat{\bm{Y}}^{u,d} $,
results in a scenario that can easily be
accessed and excluded directly at the LHC. In fact, such an alignment would lead to (approximately) the same relative shift in the down 
and bottom Yukawa couplings. Our method thus offers a complementary indirect access to the diagonal Yukawa coupling of light 
quarks, given that the LHC finds no large deviation in the bottom Yukawa. In that case, which excludes large contributions from a
potentially flavor-aligned deviation in the Yukawa couplings (evading FCNCs), our approach has a high power in constraining 
$Y_d$. In the following we will indeed assume the absence of sizable corrections in $Y_b$ of $> {\cal O}(20 \%)$ and then our 
predictions can be considered generic.

We thus start from arbitrary complex Yukawa matrices $\hat{\bm{Y}}^{u,d}$ in the flavor basis and introduce a misalignment
with the mass matrices $\hat{\bm{M}}^{u,d}$ via the coefficients of the anarchic $D=6$ operators $\bm{C}_{u,d}$, see (\ref{eq:D6})
and (\ref{eq:LM}). The only restriction we impose on $\hat{\bm{M}}^{u,d}$ is that they reproduce the correct mass eigenvalues and 
the CKM matrix after
 diagonalization, {\it i.e.}, 
	\begin{equation}
	\bm{U}_L^d = \bm{U}_L^u\, \bm{V}_{\rm CKM},
	\end{equation} 
where 
	\begin{equation}
	\hat{\bm{M}}^d = \bm{U}_L^d\,{\rm diag}(m_d,m_s,m_b)\,\bm{U}_R^{d\, \dagger}\, , \quad
	\hat{\bm{M}}^u = \bm{U}_L^u \,{\rm diag}(m_u,m_c,m_t)\, \bm{U}_R^{u\, \dagger}.
	\end{equation}
The Higgs-coupling matrices  in the physical basis ${\bm{Y}}^{u,d}$ are then obtained via
	\begin{equation}
 	{\bm{Y}}^d =  \bm{U}_L^{d\, \dagger} \hat{\bm{Y}}^d  \bm{U}_R^d
	\end{equation}
and
	\begin{equation}
 	{\bm{Y}}^u =  \bm{U}_L^{u\, \dagger}  \hat{\bm{Y}}^u \bm{U}_R^u
	=  \bm{V}_{\rm CKM}  \bm{U}_L^{d\, \dagger}  \hat{\bm{Y}}^u  \bm{U}_R^u\,.
	\end{equation}
It is just the diagonal (1,1) entries of these Higgs couplings $ Y_{u,d} \equiv (\bm{Y}^{u,d})_{11}$, that we want 
to constrain from experimental input in the following.

It is important to note that in general one basis for the mass matrices is as good as the other, and only the misalignment 
between $\hat{\bm{M}}^u$ and $\hat{\bm{M}}^d$ is physically observable through the CKM matrix. In consequence, it seems 
reasonable to assume the most general modifications of the SM values for the Yukawa couplings in the 
original basis. In particular, in an anarchic approach, it would be unnatural for $\bm{C}_{u,d}$ to be diagonal in the same 
basis as $\hat{{\bm{Y}}}^{u,d}_{\rm SM}$. In the following section we will present our numerical results. We will first employ
arbitrary complex numbers in the range $v/\sqrt 2 |(\bm{C}_{u,d})_{ij}|= \left[0,5\right]$\,MeV, evaluated at the low scale of the 
experiments, and demonstrate the correlation between $Y_{u,d}$ and FCNCs. To show that our obtained constraints 
on $Y_{u,d}$  from flavor-physics experiments are to good approximation insensitive to our assumptions for the Wilson 
coefficients, we will study two further scenarios. In the first we consider a larger scale $v/\sqrt 2 |(\bm{C}_{u,d})_{ij}|=
\left[0,0.1\right]$\,GeV and in the second we only put $v/\sqrt 2|(\bm{C}_{u,d})_{1,1}|= \left[0,5\right]$\,MeV, and the 
other entries vanishing, in order to demonstrate that we do not feed in special off-diagonal transitions ``by hand''.

\section{Results}

	\begin{figure}[!t]
	\begin{center}
	\includegraphics[height=2.25in]{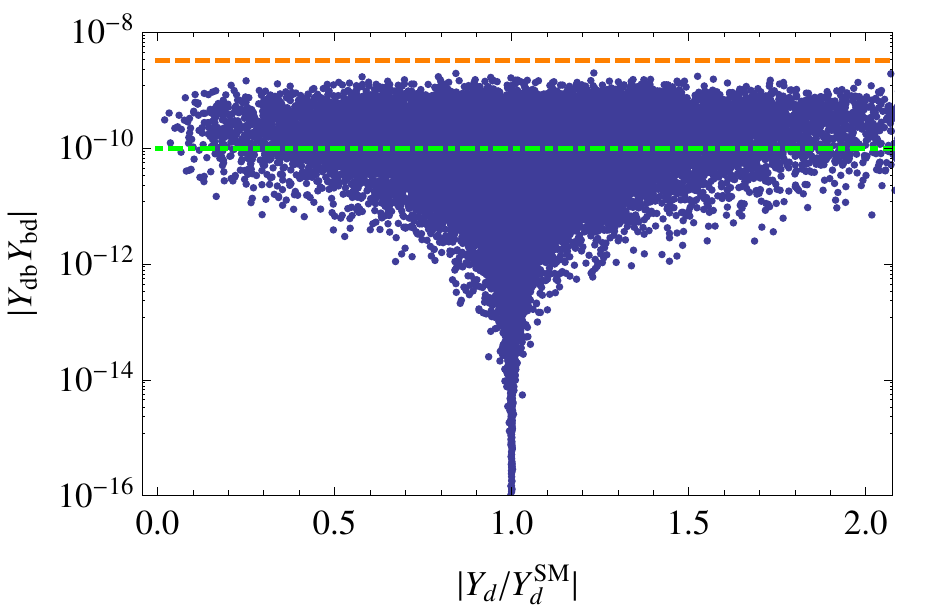}
	\parbox{15.5cm}{\caption{\label{fig:dbbd}
	 Predictions for the correlation between the off-diagonal transition $|(\bm{Y}^{d})_{13}(\bm{Y}^{d})_{31}|$
	and $|Y_d/Y_d^{\rm SM}|$. The current experimental constraint from $B_d^0$ oscillations is given as an orange dashed line, 
	while a potential experimental improvement by around an order of magnitude is depicted as a green dot-dashed line. See text for 
	details.}}
	\end{center}
	\end{figure}
In Figure \ref{fig:dbbd}, we show the flavor changing combination of couplings $|(\bm{Y}^{d})_{13}(\bm{Y}^{d})_{31}|$ in dependence 
on the absolute value of the ratio of the physical down-quark Yukawa coupling $Y_d$ and its value in the SM (with $\bm{C}_{u,d}=
\bm{0}$), $Y_d^{\rm SM}=m_d \sqrt 2/v$. Here and in the following we scan the parameterspace uniformly in the complex 
plane within $|(\bm{C}_{u,d})_{ij}|= \left[0,5\right]$\,MeV, fixing ${\bm{Y}}^{u,d}_{\rm SM}$ in an agnostic way that reproduces
the correct quark masses and mixings. We note that the crucial lower contour of the scatter plots is to good approximation 
independent of continuous changes of the range of the paramters, see below. It should nevertheless be stressed that the analysis 
is meant to examine the general picture and does not take into account all fine-tuned parameterpoints that might be possible. 
We also give the experimental upper limit on the corresponding off-diagonal Yukawa couplings from flavor physics ($B_d^0$ oscillations) 
as presented in \cite{Harnik:2012pb}, $|(\bm{Y}^{d})_{13}(\bm{Y}^{d})_{31}|_{\rm exp}<3.3\times 10^{-9}$, as an orange dashed 
line (see \cite{Bona:2007vi} for the corresponding measurements). 

One can clearly see that already deviations of the order of 
$10\%$ in $Y_b$   lead generically to non-negligible flavor changing effects of $|(\bm{Y}^{d})_{13} (\bm{Y}^{d})_{31}|>10^{-12}$ which are 
however not yet excluded by experiment. A vanishing $Y_d$ would on the other hand result in $|(\bm{Y}^{d})_{13}(\bm{Y}^{d})_{31}|>10^{-10}$. 
So while with current data on $B_d^0$ oscillations, one can not yet discard the $Y_d=0$ hypothesis, a modest experimental improvement in the limit 
of around one order of magnitude to $|(\bm{Y}^{d})_{13}(\bm{Y}^{d})_{31}|_{\rm exp}<10^{-10}$, depicted by the green dot-dashed line, 
could already strongly disfavor the $Y_d=0$ hypothesis. However, we conclude that current limits on $B_d^0$ oscillations are not yet capable of 
constraining $Y_d$ in an interesting range of modest deviations from the SM.

	\begin{figure}[!t]
	\begin{center}
	\includegraphics[height=1.89in]{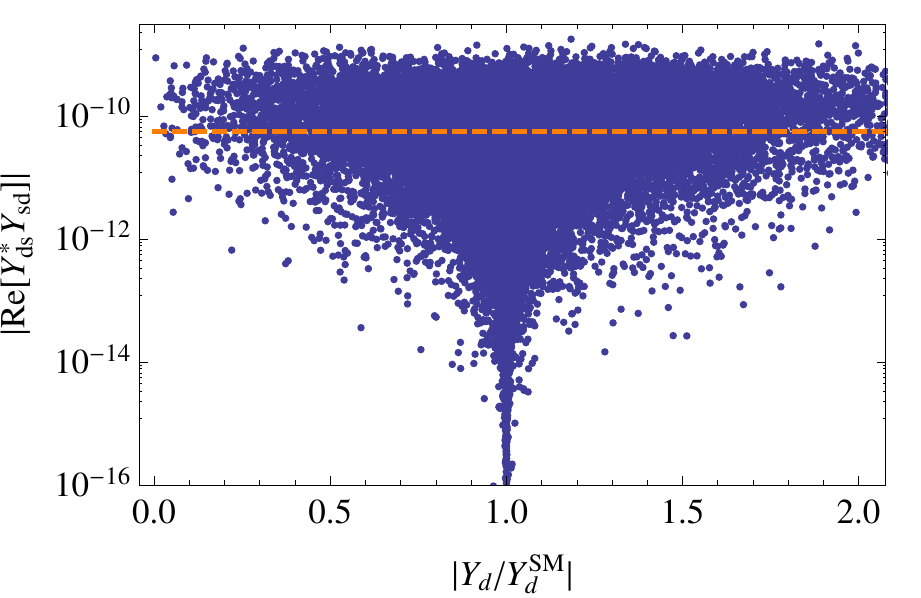}\ \ \includegraphics[height=1.89in]{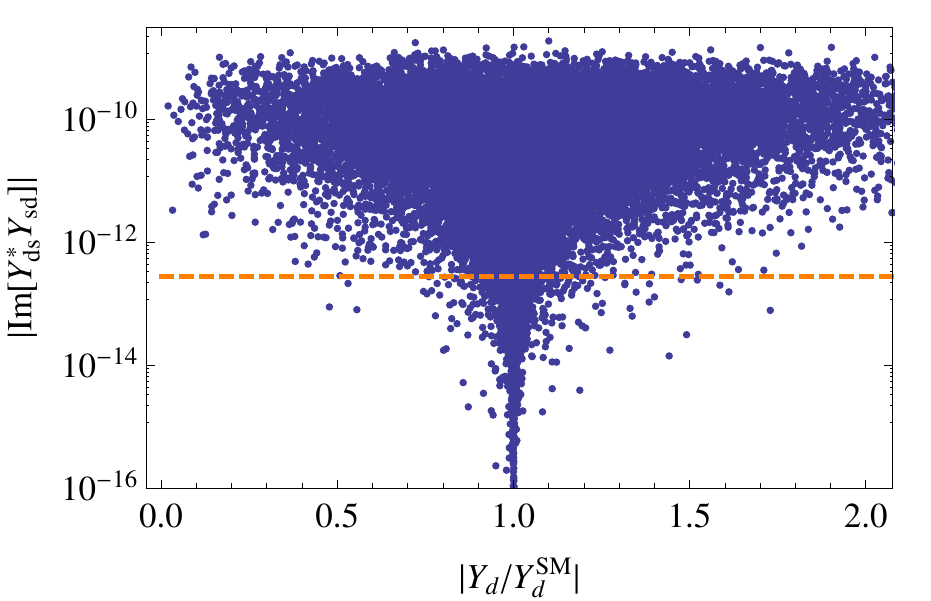}
	\parbox{15.5cm}{\caption{\label{fig:K}
	 Predictions for the correlation between $|{\rm Re}[(\bm{Y}^{d\, \ast})_{12}(\bm{Y}^{d})_{21}]|$, $|{\rm Im}
	[(\bm{Y}^{d\, \ast})_{12}(\bm{Y}^{d})_{21}]|$ and $Y_d$. The current experimental constraints from $K^0$ oscillations 
	are given as the orange dashed lines. See text for details.}}
	\end{center}
	\end{figure}
This situation changes when we take into account Kaon physics. In the left (right) panel of Figure \ref{fig:K}, we show the magnitude of
the real part (imaginary part) of the squared $ds$ transitions $|{\rm Re}\left[(\bm{Y}^{d\, \ast})_{12}(\bm{Y}^{d})_{21}\right]|$, $|{\rm Im}
\left[(\bm{Y}^{d\, \ast})_{12}(\bm{Y}^{d})_{21}\right]|$ versus the absolute value of the ratio of the physical down-quark Yukawa coupling 
$Y_d$ over $Y_d^{\rm SM}$. We also give the experimental upper limit on the corresponding off-diagonal Yukawa couplings 
from Kaon physics, $|{\rm Re}\left[(\bm{Y}^{d\, \ast})_{12}(\bm{Y}^{d})_{21}\right]|_{\rm exp}<5.6\times 10^{-11}$,
$|{\rm Im}\left[(\bm{Y}^{d\, \ast})_{12}(\bm{Y}^{d})_{21}\right]|_{\rm exp}<2.8\times 10^{-13}$ 
\cite{Harnik:2012pb}, as an orange dashed line. We note that we make no special assumption on the phases present in (\ref{eq:D6}).  
It is evident from the plots that, while the constraint on the real part of the couplings has only a marginal constraining power so far, the 
limit on the imaginary part already allows the conservative estimate
\begin{equation}
0.4<|Y_d/Y_d^{\rm SM}|<1.7\,.
\end{equation}
While this is not a limit that one can not avoid via fine-tuning the structure of the Yukawa matrices, it however provides us with a rather
stringent range where we expect $Y_d$ to lie, given the FCNC data. Let us stress that no other measurement so far exhibits a comparable 
power in unveiling information on $Y_d$. Future improvements in FCNC measurements are expected to provide even tighter constraints 
on $Y_d$.

\begin{figure}[!t]
	\begin{center}
	\includegraphics[height=2.22 in]{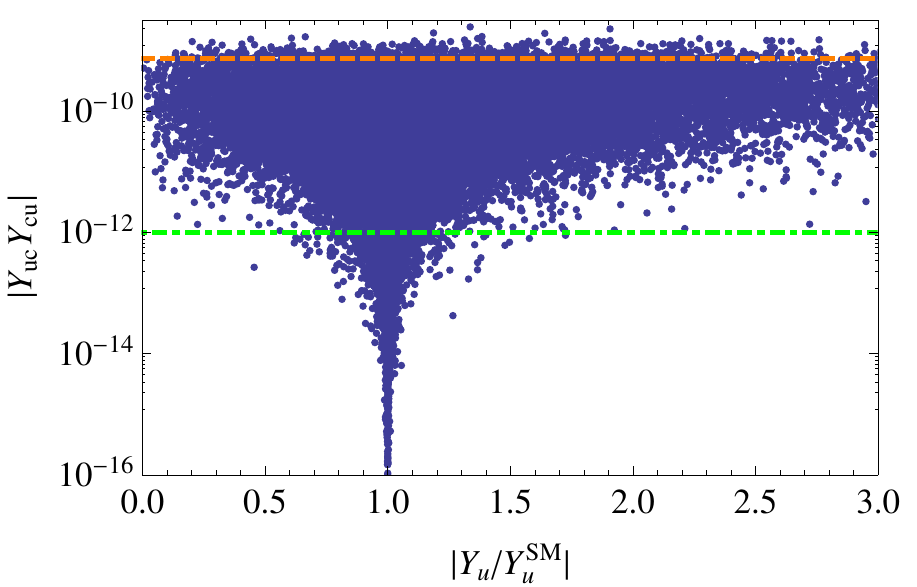}
	\parbox{15.5cm}{\caption{\label{fig:D}
	 Predictions for the correlation between the off-diagonal transition $|(\bm{Y}^{u})_{12}(\bm{Y}^{u})_{21}|$
	and $|Y_u/Y_u^{\rm SM}|$. The current experimental constraint from $D_d^0$ oscillations is given as an orange dashed line. 
	See text for details.}}
	\end{center}
	\end{figure}

Turning our attention to the up-quark sector, we note that the most promising limits from $D^0$ oscillations 
$|(\bm{Y}^{u})_{12}(\bm{Y}^{u})_{21}|_{\rm exp}<7.5 \times 10^{-10}$ \cite{Harnik:2012pb} are not yet providing strong constraints 
on $Y_u$. This is visualized in Figure \ref{fig:D}, where we show $|(\bm{Y}^{u})_{12}(\bm{Y}^{u})_{21}|$ versus the absolute value of the ratio 
of the physical up-quark Yukawa coupling $Y_u$ over $Y_u^{\rm SM}$. The experimental limit is again indicated by the orange dashed line. 
Here, an improvement of 2-3 orders of magnitude, indicated by the green dot-dashed line, is necessary in order to derive stringent 
constraints on $Y_u$.

\begin{figure}[!h]
	\begin{center}
	\includegraphics[height=3.85 in]{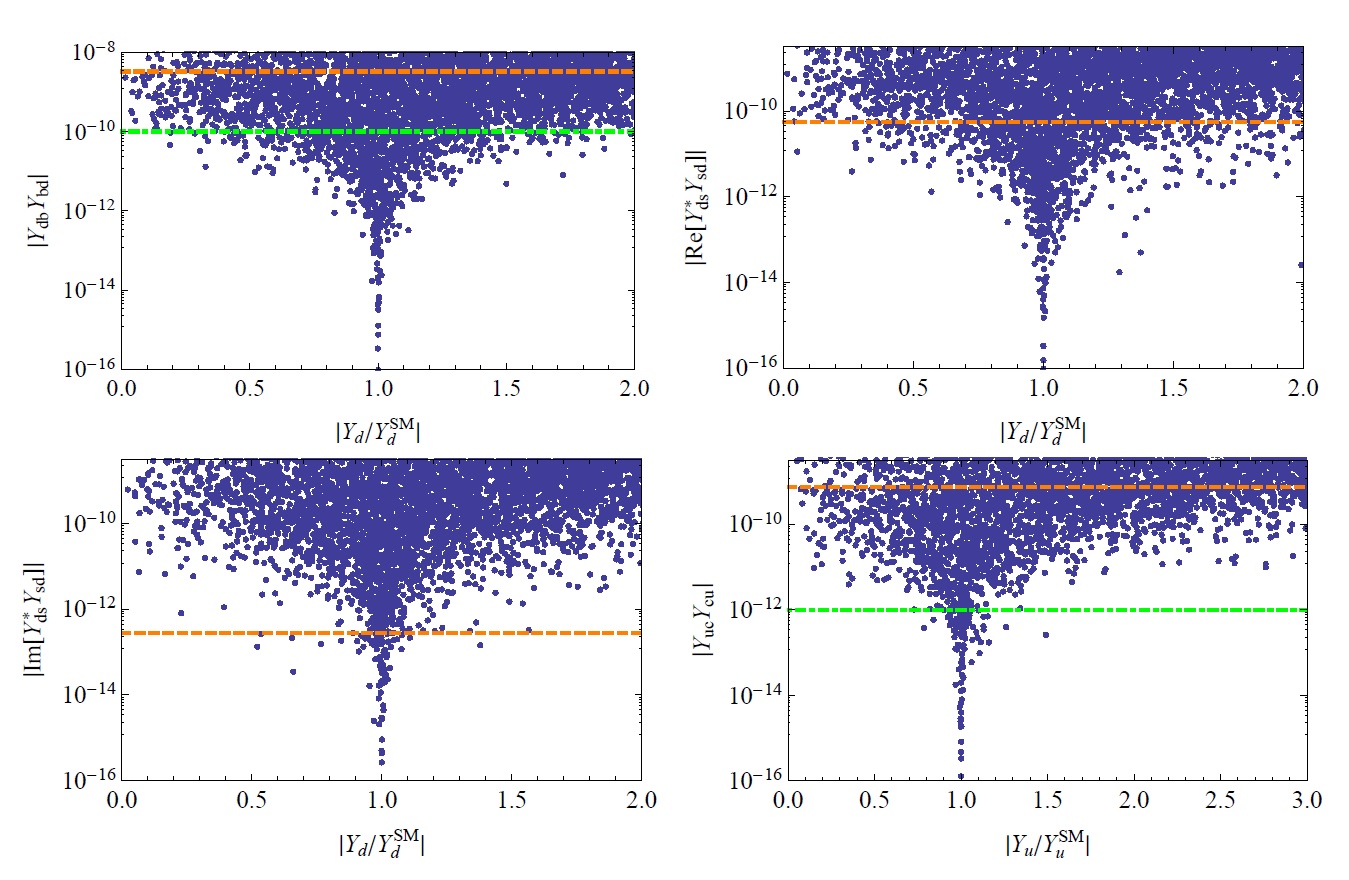}
	\parbox{15.5cm}{\caption{\label{fig:2}
	Same plots as before, now with $v/\sqrt 2 |(\bm{C}_{u,d})_{ij}|= \left[0,0.1\right]$\,GeV, see text for details.}}
	\end{center}
	\end{figure}

\begin{figure}[!h]
	\begin{center}	
	\includegraphics[height=3.85 in]{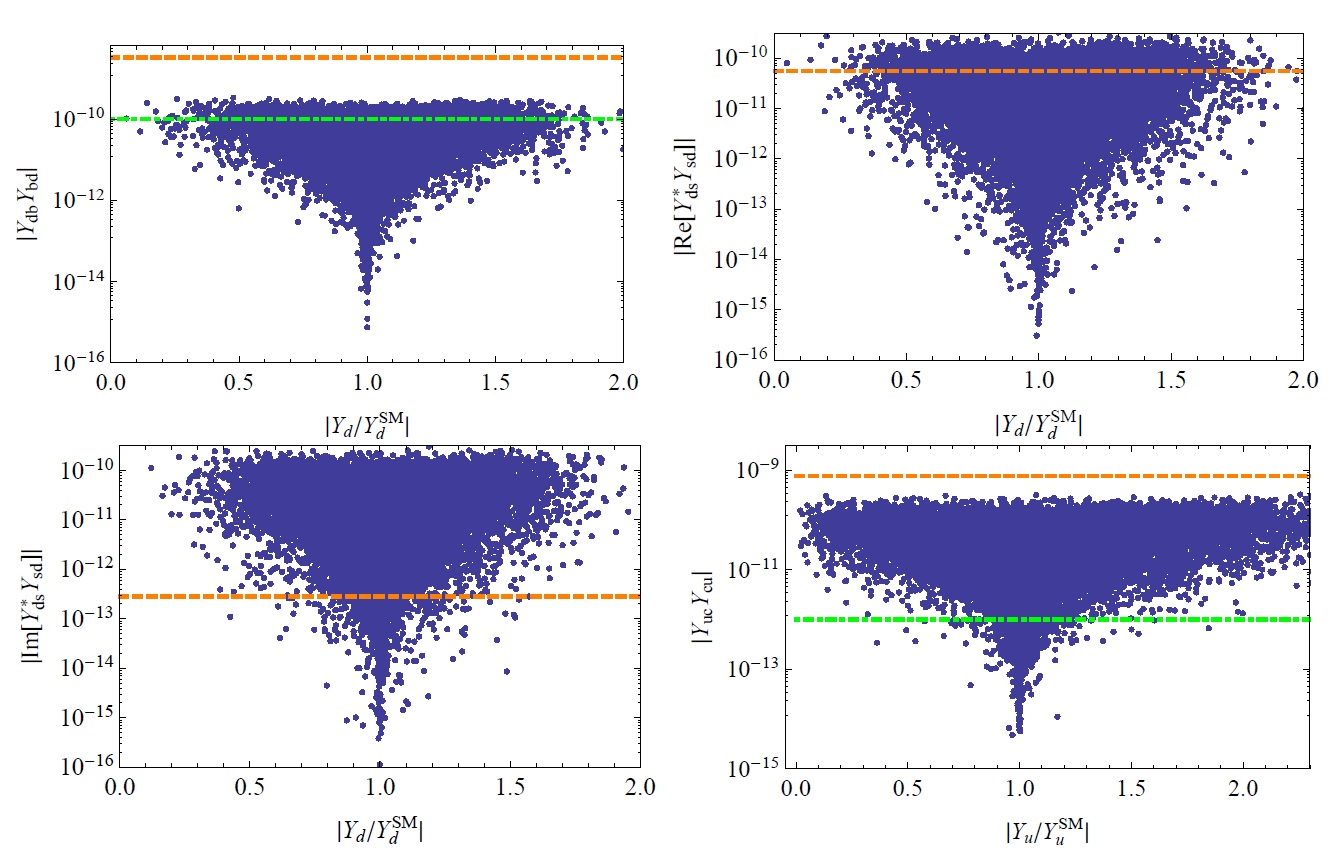}
	\parbox{15.5cm}{\caption{\label{fig:3}
	Same plots as before, now with $v/\sqrt 2|(\bm{C}_{u,d})_{1,1}|= \left[0,5\right]$\,MeV, see text for details.}}
	\end{center}
	\end{figure}

Finally, to show that our findings are robust with respect to continuous deformations of our setup, we will now study two such possible
modifications \footnote{Note that the drastic scenario of a full generation of the mass matrices via a different source than the Higgs boson, 
which would lead to a vanishing $Y_{u,d}$ without necessarily introducing new FCNCs, invalidating our approach, is already highly disfavored 
from Higgs physics at the LHC.}.
First, we will raise the scale of the operators by considering $v/\sqrt 2 |(\bm{C}_{u,d})_{ij}|= \left[0,0.1\right]$\,GeV. Then, in order to show
that we do are not putting in off-diagonal transitions artificially in an ad hoc way, we are considering $v/\sqrt 2
|(\bm{C}_{u,d})_{1,1}|= \left[0,5\right]$\,MeV, and the other entries vanishing.
The results are given in Figures \ref{fig:2} and \ref{fig:3}, respectively, which contain all plots shown so far with an adjusted parameterspace, 
as discussed above. The plots confirm that the lower contours, which provide the important connection between limits on FCNCs and the 
physical Yukawa couplings $Y_{u,d}$ are rather independent of the particular assumptions on the operators.

\section{Conclusions}
\label{sec:conclusions}

We have shown how negative search results for FCNCs transitions can be turned into valuable constraints on the first generation Yukawa
couplings. While it seems hopeless to get direct information on these couplings from Higgs physics in the near future, given we find no large 
deviations in $Y_b$, our method provides the estimate 
$0.4<|Y_d/Y_d^{\rm SM}|<1.7$. To obtain statements of similar quality for the up-quark sector, some improvements in experimental limits
on FCNCs are required.

\paragraph{Acknowledgements}

I am grateful to Babis Anastasiou, Adam Falkowski, and Uli Haisch for useful comments. 
The research of the author is supported by the Swiss National Foundation under contract SNF 200021-143781.

\end{document}